\documentclass[11pt,twoside]{article}

\usepackage{asp2014}

\aspSuppressVolSlug
\resetcounters

\begin{document}

\title{The Future of astronomical archives: reaching out to and engaging broader communities}

\author{Raffaele D'Abrusco,$^1$, Glenn Becker,$^1$, Michael McCollough,$^1$, Sinh Thong,$^1$, David VanStone,$^1$, 
Sherry Winkelman,$^1$}

\affil{$^1$Center for Astrophysics | Harvard \& Smithsonian, Cambridge (MA), USA; \email{rdabrusc@cfa.harvard.edu}}

\paperauthor{Raffaele D'Abrusco}{rdabrusc@cfa.harvard.edu}{0000-0003-3073-0605}{Center for Astrophysics | Harvard \& 
Smithsonian}{High Energy Astrophysics Division}{Cambridge}{MA}{02138}{USA}
\paperauthor{Glenn Becker}{gbecker@cfa.harvard.edu}{}{Center for Astrophysics | Harvard \& 
Smithsonian}{High Energy Astrophysics Division}{Cambridge}{MA}{02138}{USA}
\paperauthor{Michael McCollough}{mmccollough@cfa.harvard.edu}{}{Center for Astrophysics | Harvard \& 
Smithsonian}{High Energy Astrophysics Division}{Cambridge}{MA}{02138}{USA}
\paperauthor{Sinh Thong}{sthong@cfa.harvard.edu}{}{Center for Astrophysics | Harvard \& 
Smithsonian}{High Energy Astrophysics Division}{Cambridge}{MA}{02138}{USA}
\paperauthor{David VanStone}{dvanstone@cfa.harvard.edu}{}{Center for Astrophysics | Harvard \& 
Smithsonian}{High Energy Astrophysics Division}{Cambridge}{MA}{02138}{USA}
\paperauthor{Sherry Winkelman}{swinkelman@cfa.harvard.edu}{}{Center for Astrophysics | Harvard \& 
Smithsonian}{High Energy Astrophysics Division}{Cambridge}{MA}{02138}{USA}



\begin{abstract}
The importance of archival science increases significantly for astrophysical observatories as they mature 
and their archive holdings grow in size and complexity. Further enhancing the science return of archival 
data requires engaging a larger audience than the mission reference community, mostly because of the growth 
of interest in multi-wavelength and transient/time variability research. Such a goal, though, can be 
difficult to achieve. In this paper I will describe a different approach to this question that, 
while minimizing technological 
friction and leveraging existing services, makes archival observations more accessible 
and increases our capability to proactively engage astronomers on potentially interesting archival 
records. Inspired by this strategy, the Chandra Data Archive team is working on two specific experimental 
projects that will hopefully demonstrate their potential while contributing 
to the maximization of the scientific return of the Chandra mission. 
\end{abstract}



\vspace{-1cm}
\section{Introduction}

Archive-based science becomes more and more important as astronomical missions mature and 
archive holdings grow. Increased usage of public archival observations in the literature 
has been observed for more than a decade for all NASA Great 
Observatories~\citep{white2009,peek2019} (Figure~\ref{fig:mission}). The {\it Chandra} archive 
makes no exception, as shown in Figure~\ref{fig:agingplot}, but there are some evidences 
that re-usage of archival observations is plateauing. Further enhancing the 
scientific return of archival data will require development of tools and services to engage a 
larger audience than the mission reference community. Such goals, though, can be cost-prohibitive and 
technically challenging if pursued through the continuous expansion of 
existing tools and interfaces that were designed to cater to the needs and requirements of the mission 
core community.

\articlefigure[scale=0.6]{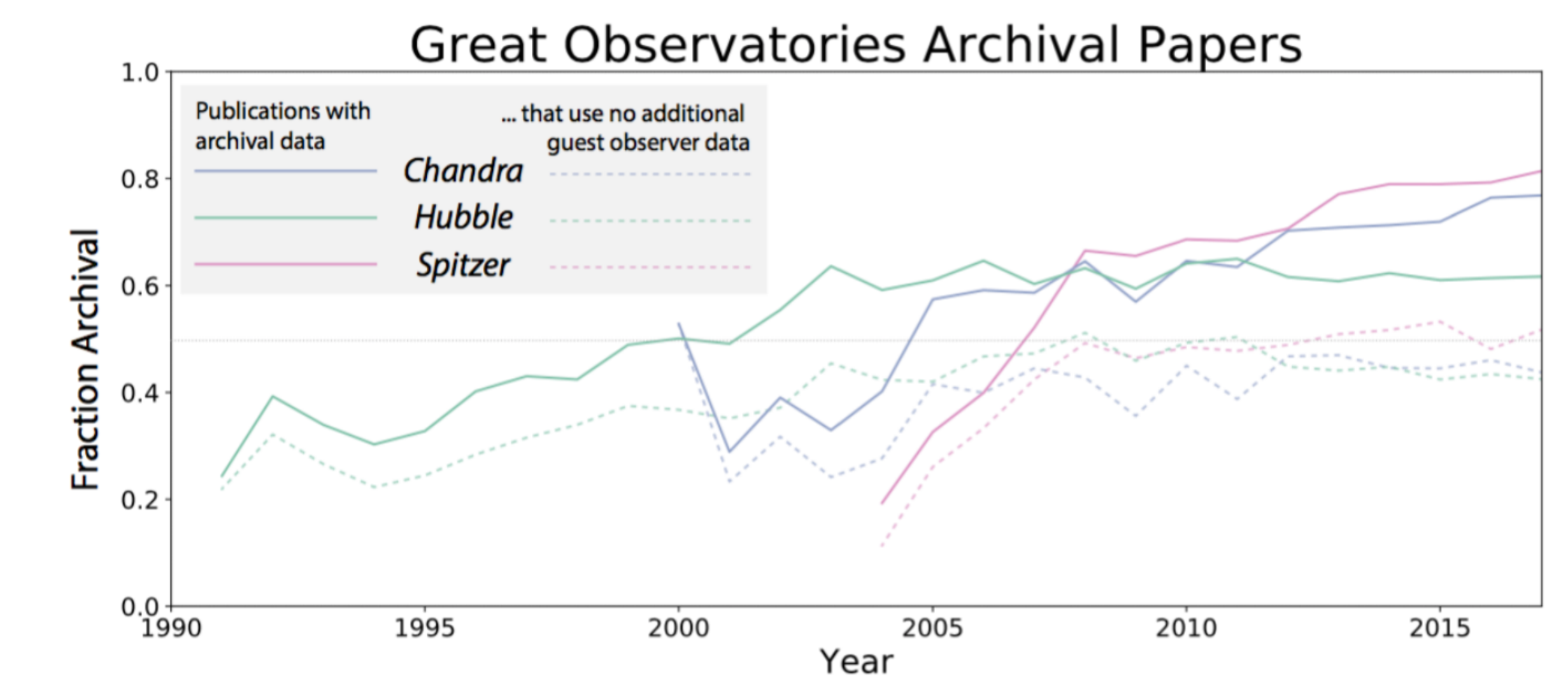}
{fig:mission}
{Fraction of papers for {\it Chandra}, the {\it Hubble Space Telescope} and {\it Spitzer} 
missions using archival data exclusively or in combination with GO data, as a function 
of time~\cite{peek2019}.}

\articlefigure[scale=0.5]{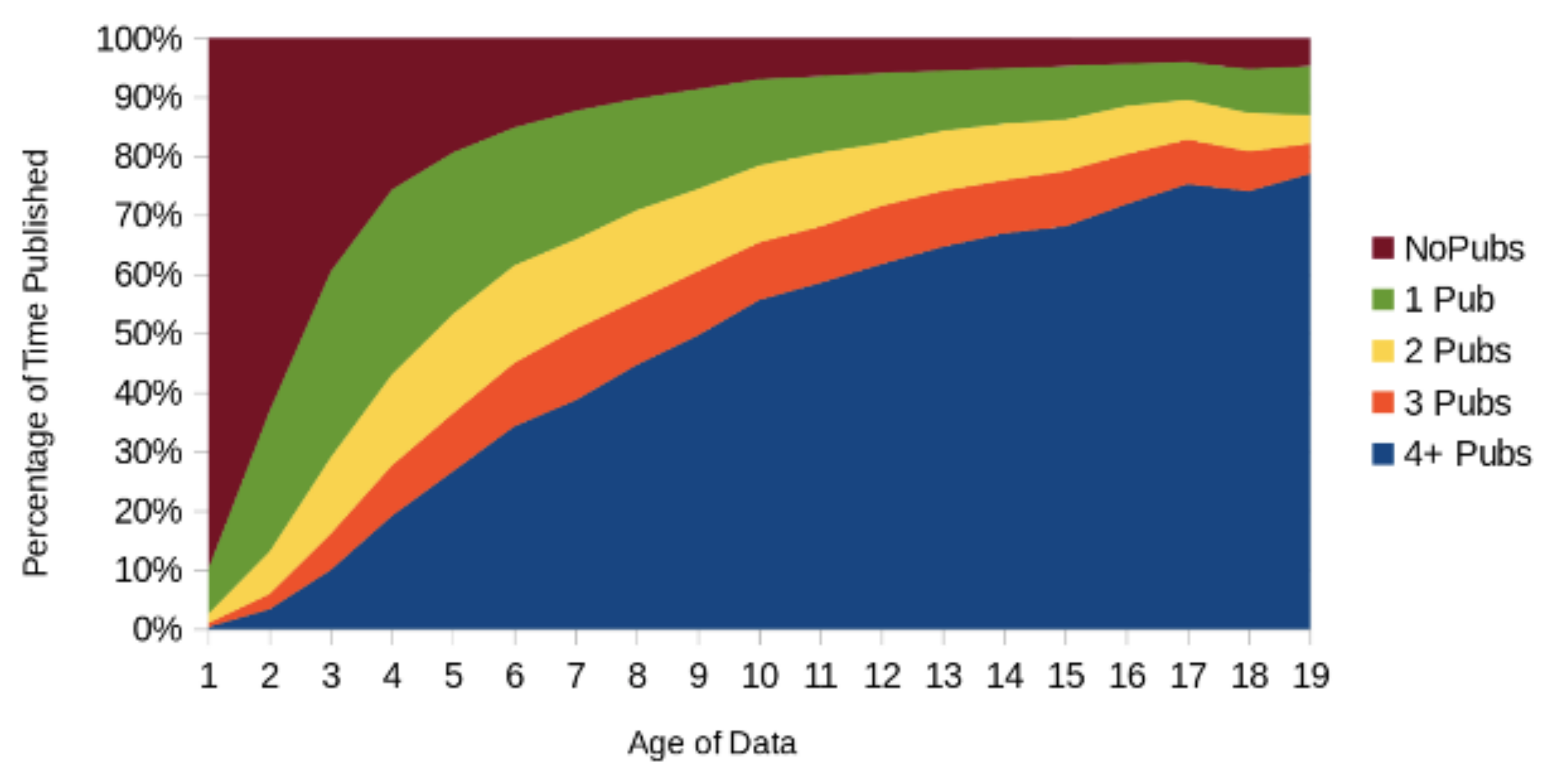}
{fig:agingplot}
{Fraction of the total exposure of Chandra public archival observations 
published in scientific literature as a function of the time spent in the 
archive. Different colors show fractions of time published more than once.}

\noindent Sustaining this growth in the medium/long term within realistic budgetary 
and technical constraints will be challenging for several 
different reasons, including:

\begin{itemize}
    \item Potential saturation of the core archival science opportunities.
    \item Degradation of the utility of public interfaces, that cannot evolve as quickly as community needs.
    \item Challenge and cost of using a unique technology to cater to well-established and new use cases. 
    \item An evolving and expanding science focus can become a moving target (transients, 
    multi-wavelength \& multi-messenger research, etc.)
\end{itemize}

\noindent So, the question is: how can archives try to sustain and push archival science towards
unexplored territories without relying on new toolkits and resources that might not be
available? One of the possible approaches is 
to develop smart methods to contact and engage a wider community than the reference one, 
stimulating new interest in the data 
available in the archive and ''out of the box``, innovative scientific applications. 
In this paper, we suggest that this goal can be accomplished by leveraging the large 
body of connections between archival observations with the literature, and taking 
advantage of external data collections to enrich its own holdings.  

\section{Sustaining growth of the CDA}

The Chandra Data Archive (CDA) is working on two different projects to reach out to a larger 
audience than the core Chandra community and generate additional, original interest in 
its holdings. This experimental approach leverages both the valuable internal collections of 
data/literature links accumulated over the first 20 years of operations of CXO in the 
Chandra bibliography~\citep{winkelman2018a,winkelman2018b}, and the 
existence of heterogeneous astronomical knowledge bases that can be used to enhance the 
knowledge content of the archive.

\subsection{...from within...}

By taking advantage of the comprehensive collection of bibliographic linkages between Chandra observations 
and the scientific and technical literature using them, we have defined multiple classes 
of interesting ``archival events'' based on public observations or {\it Chandra}-related 
publications. The categories of events and the definition of quantitative thresholds needed to 
select events are entirely data-driven, in order to guarantee objectivity, and relieve humans 
from the responsibility of the choice. 

\articlefigure[angle=270,scale=0.38]{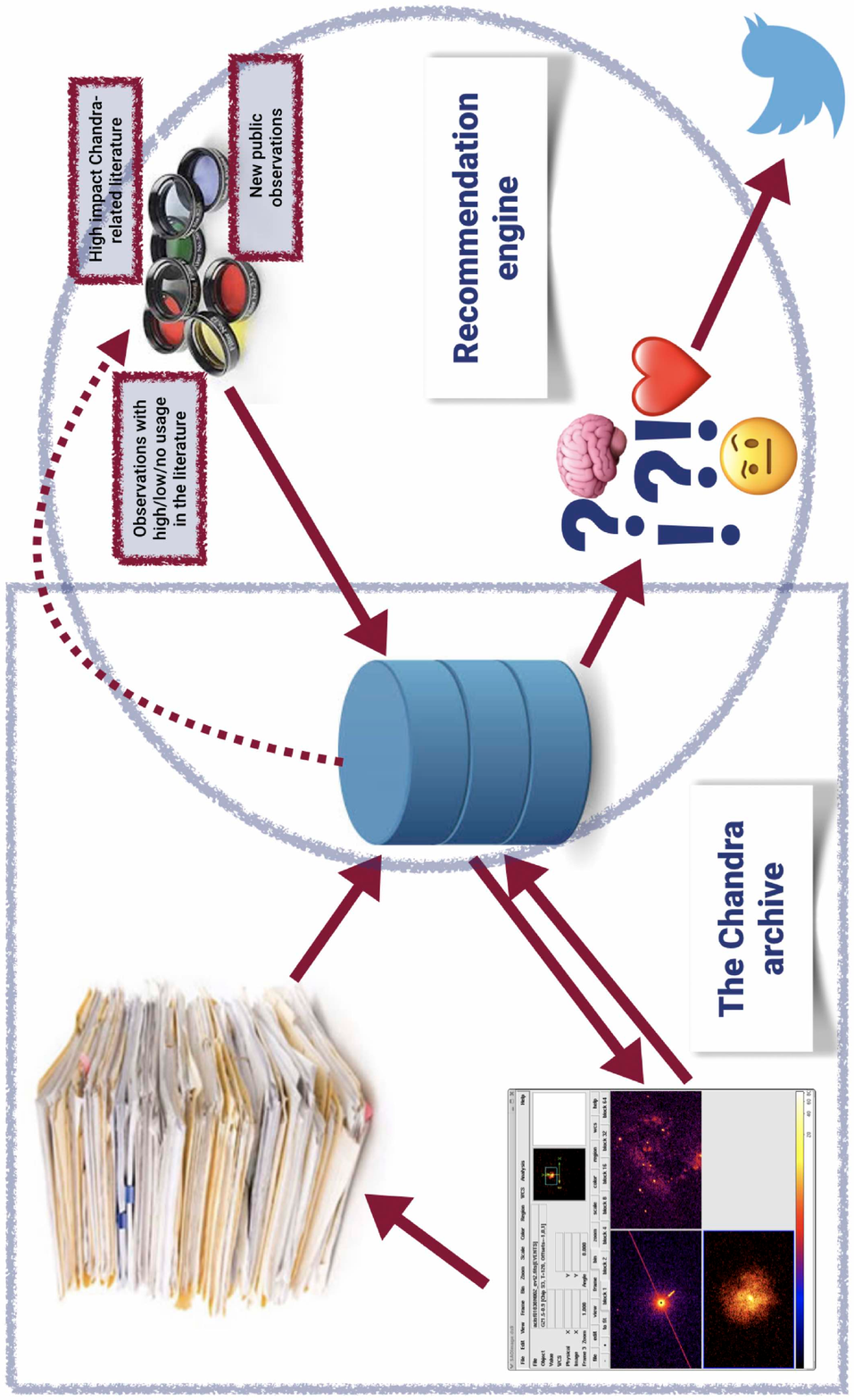}
{fig:diagram}
{Schematic representation of the Chandra Data Archive and its ``archival recommendation engine''
in the context of the automatic extraction of interesting archival events for the Twitter CDA
program.}

\noindent The ``events'' are automatically extracted from the archive by a suite of customized DB queries and 
scripts developed for the purpose, that we have called the ``archival recommendation engine''. 
In the future, archival events will be daily evaluated by CDA staff for publication through 
Twitter~\citep{becker2018}. Figure~\ref{fig:diagram} shows a diagrammatic representation of how
the recommendation engine harnesses the Chandra bibliographic archive: the bibliographic database
is searched for potential interesting events
fitting the pre-defined categories, that are then passed with all the salient information to the 
operators in a human-friendly format. At this point, all necessary ingredients are available and the 
creative process that will lead to the publication of the Tweet describing the event can start. 

\subsection{...and the outside}

In the case of most archival interfaces, searches are limited to metadata of the observations (space/time 
location, size, target, instrumental parameters, duration, mode). While there has been discussion about 
how to bring natural language searches into the astronomical research, progress has been relatively slow, and 
a general mechanism to allow more useful questions like the one reported below is still missing:\newline

 ``{\it What are the ACIS observations containing Seyfert 2 galaxies with redshift between 0.5 and 0.8 and 
 located less than 0.15 degrees from the coordinates of the pointings of the Chandra observations with 
 exposure (at the location of the galaxies) longer than 25 ks?}''\newline

The Chandra archive does not know what Seyfert 2 galaxies and redshift are. In CDA, we plan to address 
this issue by harnessing an external service that collects, combines and curates heterogeneous astronomical 
information, namely the SIMBAD Astronomical Database~\citep{wenger2000} managed by the Centre de Donn\'ees 
astronomiques de Strasbourg (CDS).
We will annotate the footprint of all Chandra public observations and provide an additional layer
to the current public interface that will make these type of questions answerable.

\section{Conclusions}

Simple recipes can be found to enhance the scientific impact of archival data leveraging 
internal and external resources to increase the astronomical value of the data and raising interest
for its re-usage. In this paper, I have described two ongoing projects at the CDA that 
that will ensure a proactive role for single mission archives in the future.

\acknowledgements This work has been supported by NASA under contract NAS 8-03060 to the 
Smithsonian Astrophysical Observatory for operation of the Chandra X-ray Center.


\bibliography{biblio_adass}  


\end{document}